\documentclass[epj]{svjour}

\usepackage{graphicx}
\usepackage{dcolumn}
\usepackage{bm}

\begin{document}

\title{Communication activity in social networks: 
growth and correlations}

\author{
Diego Rybski\inst{1,2} 
\and 
Sergey V. Buldyrev\inst{3} \and
Shlomo Havlin\inst{4} \and 
Fredrik Liljeros\inst{5} \and
Hern\'an A. Makse\inst{1}}
\institute{
Levich Institute and Physics Department, 
City College of New York, 
New York, NY 10031, USA
\and 
Potsdam Institute for Climate Impact Research (PIK),
P.O. Box 60 12 03,
14412 Potsdam,
Germany
\and
Department of Physics, 
Yeshiva University, 
New York, NY 10033, USA
\and
Department of Physics, 
Bar-Ilan University, 
Ramat-Gan 52900, Israel
\and
Department of Sociology, 
Stockholm University, 
S-10691 Stockholm, Sweden
}

\date{\today\enspace[version 12]}

\abstract{
We investigate the timing of messages sent in two online communities 
with respect to growth fluctuations and long-term correlations. 
We find that the timing of sending and receiving messages comprises
pronounced long-term persistence.  Considering the activity of the
community members as growing entities, i.e. the cumulative number of
messages sent (or received) by the individuals, we identify
non-trivial scaling in the growth fluctuations which we relate to the
long-term correlations.  We find a connection between the scaling
exponents of the growth and the long-term correlations which is
supported by numerical simulations based on peaks over threshold.
In addition, we find that the activity on directed links
between pairs of members exhibits long-term correlations,
indicating that communication activity with the most liked partners
may be responsible for the long-term persistence in the timing of
messages.  Finally, we show that the number of messages, $M$, and the
number of communication partners, $K$, of the individual members are
correlated following a power-law, $K\sim M^\lambda$, with exponent
$\lambda\approx 3/4$.
}

\maketitle

\setcounter{tocdepth}{4}
\tableofcontents

\section{Introduction}
\label{sec:intro}

Seeking for simple laws and regularities in human activity,
researchers belonging to various disciplines aim to study social
phenomena by describing them with methods from natural sciences.
Since communication plays a predominant role in social systems, it is
desired to obtain better insight into the nature of communication
patterns -- and therefore to understand both, communication itself
and the social systems. Although it is clear that communication
is related to the embedment in social networks, the actual dynamical
processes are still poorly understood.

Studying economic data, surprising growth patterns have been
identified \cite{StanleyABHLMSS1996}, which seem to be abundant in
systems with growth-like features
\cite{CanningALMS1998,PlerouAGMS1999,LiljerosF2003,MatiaALMS2005,PicoliM2008,RozenfeldRABSM2008,WangYHS2009}.
Considering the units of a system of interest and calculating their
logarithmic growth rates between two time steps, it was found that the
standard deviation of the growth rates decays as a power-law with
the initial size \cite{StanleyABHLMSS1996}.
This finding represents a violation of Gibrat's law 
\cite{GibratR1931,SuttonJ1997,MitzenmacherM2004} 
stating that the average and the standard deviation of the growth rate of a
given economic indicator are constant and independent of the specific
indicator value, see also \cite{RozenfeldRABSM2008}.

In a recent study \cite{RybskiBHLM2009} we have found
several scaling laws characterizing the communication activity in
online social networks.  We found the existence of long-term
correlations in human activity of sending messages to other members
in the social network. The long-term persistence is related to the
fluctuations in the growth properties of the social network as
measured by the cumulative number of message sent by the members.
The present paper expands this previous work by studying the messages
sent in two online social networks with respect to the following
properties.  First, we extend the results obtained in
\cite{RybskiBHLM2009}, revealing the analogue correlations in the
timing of receiving messages.
Furthermore, we analyze the temporal correlations of the activity on
directed links, i.e. between pairs of members, and find almost
identical results as on the level of the single members.

Second, in line with \cite{RybskiBHLM2009} we study the growth of the
cumulative communication activity of the members in terms of the
cumulative numbers of messages sent and received.  In
\cite{RybskiBHLM2009} we have shown that the standard deviation of the
growth rates of the cumulative number of messages sent by individuals
depend on the 'size' of the member (defined as the cumulative numbers
of messages) following a power-law with exponent $\beta\approx 0.2$,
significantly different from the random exponent $\beta_{\rm
  rnd}=1/2$, indicating nontrivial fluctuations and persistence in the
human communication activity in the social networks. Here we further
study the distribution of the logarithmic growth rates and find
exponential decays similarly to those encountered in econophysics
\cite{StanleyABHLMSS1996}.

Third, in order to understand the relation between the long-term
correlations and growth fluctuations, we propose a simulation approach 
based on peaks over threshold modeling.  Using artificially generated
long-term correlated sequences, a message is sent when the record
exceeds a predefined threshold.  Numerically, we measure the long-term
correlations characterized by the exponent $H$ as well as growth
fluctuations characterized by $\beta$ and find that the relation
connecting both exponents proposed in \cite{RybskiBHLM2009} holds.

Fourth, we introduce a new growth rate between any pair of members
quantifying the mutual growth in the number of messages.  We find that
the corresponding growth fluctuations follow a power-law with similar
exponent as for the 'normal' growth rates.  We motivate that the
exponent might be related to cross-correlations in the activity of the
members.

Fifth, in addition to the temporal correlations, we investigate the
total number of messages sent or received and the total in- and
out-degree (i.e. the number of \emph{different} members from
which a member receives or to whom he/she sends). 
We find that the total degree and the final number of messages are
correlated following a power-law with exponent close to $0.75$.
In the case of final in- vs. out-degree, deviations from the linear
correlations are found.

Finally, we point out that there is also
a relation between our results on growth fluctuations and
long-range correlations ($\beta$ and $H$, respectively) and the
existence of power-law distributed inter-event times characterized
by the exponent $\delta$ \cite{BarabasiAL2005} leading to the
clustering and bursts in the activity of members.  This connection is
explored in a follow-up paper \cite{RybskiBHLM5a}.

Our results have important implications for the design of
communication systems. The correlations can be elaborated to better
predict information propagation, see e.g. \cite{KarsaiKPKKBS2010}.  In
addition, the characterization of fluctuations is essential for the
knowledge of uncertainty.  Our approach could be also applied in
natural systems such as in the context of protein unfolding
\cite{BrujicHWF2006}.

This paper is organized as follows. In Sec.~\ref{sec:data} we briefly
describe the data of messages sent in two online communities.
Our results are presented in Sec.~\ref{sec:analysis} which is
organized in four sub-sections -- discussing long-term correlations,
growth fluctuations, modeling, and other correlations. 
Finally, we draw our conclusions in Sec.~\ref{sec:conclusions}.

\section{Data}
\label{sec:data}

We analyze the timing of messages sent in two Internet communities
\cite{RybskiBHLM2009,GallosRLHM2011}. 
The data of the first online community (www.qx.se, QX)
\footnotemark[0]\footnotetext[0]{The study of the de-identified dating
  site network data was approved by the Regional Ethical Review board
  in Stockholm, record 2005/5:3.}
consists of over
$80,000$~members and more than $12.5$~million messages sent during
$63$~days (mid November 2005 until mid January 2006).
The data of the second online community (www.pussokram.com, POK) 
covers 492 days 
(February 2001 until June 2002) of activity with more than 
500,000 messages sent among almost 30,000 members
\cite{HolmeP2003,HolmeLEK2003,HolmeEL2004}.
This corresponds to the entire lifespan of the social network. 
Both web-sites are used for dating and general social interactions.
The QX community is used mainly by Swedish gay and lesbian while
POK was targeted to Swedish teenagers and young adults. All data
are completely anonymous, lack any message content and consist only of
the time when the messages are sent and identification numbers of the
senders and receivers.  The advantage of these data sets is that they
provide the exact time when the messages were sent -- in contrast to
similar network data sets consisting only of snapshots,
i.e. temporally aggregated social networks expressing who sent
messages to whom (see \cite{GallosRLHM2011} for a discussion).

Similarly to other online communities, the members can log in and meet
virtually. There are different ways of interacting in these
communities. Common among most of such online communities is the
possibility to choose favorites, i.e. a list of other members, that a
person somehow feels committed to. In addition, the platforms offer
the possibility to join groups and discuss with other members about
specific topics. We focus on the messages sent among the
members. These messages are similar to e-mails but have the advantage
that they are sent within a closed community where there are no
messages coming from or going outside.

From the message data one can also build networks, 
which consist of links connecting nodes. 
We consider the members as nodes and set a directed link from node 
\texttt{a} to \texttt{b} when member \texttt{a} sends at least 
one message to \texttt{b}. 
The degree, $k$, of a node is the number of other nodes it is connected to, 
i.e. the number of links it has. 
In the directed case one distinguishes between 
out-degree (number of out-going links) and 
in-degree (number of in-going links).

\section{Analysis}
\label{sec:analysis}

\subsection{Long-term correlations}
\label{subsec:resltc}

\begin{figure}
\begin{centering}
\includegraphics[width=\columnwidth]{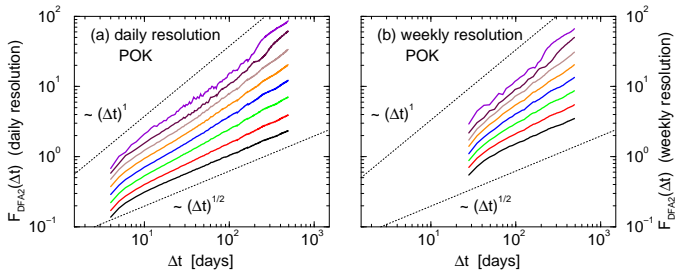}
\caption{
\label{fig:dfaaveallbind1v2}
Comparison of fluctuation functions in (a)~daily and (b)~weekly resolution of 
members sending messages in POK. 
The different curves correspond to different activity levels: 
$M=$1-2, 3-7, 8-20, 21-54, 55-148, 149-403, 404-1096, 1097-2980 
total messages (from bottom to top). 
The curves in (b) have been shifted along the $\Delta t$ axis 
to match daily resolution.
In both cases the asymptotic scaling is the same.
The dotted lines correspond to the exponents $H=1$ (top) and $H=1/2$ (bottom).
}
\end{centering}
\end{figure}
\begin{figure}
\begin{centering}
\includegraphics[width=\columnwidth]{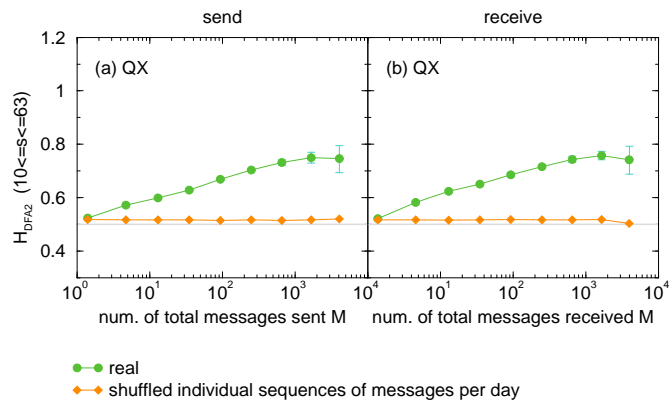}
\caption{
\label{fig:oc1dfaaveallbinfit2}
Fluctuation exponents of the communication activity (a)~sending and 
(b)~receiving messages by members of QX. 
The exponents are plotted as a function of the activity level~$M$, 
i.e. total number of messages, 
for the original data (green circles), and 
individually shuffled sequences (orange diamonds).
See also~\cite{RybskiBHLM2009}.
}
\end{centering}
\end{figure}
\begin{figure}
\begin{centering}
\includegraphics[width=\columnwidth]{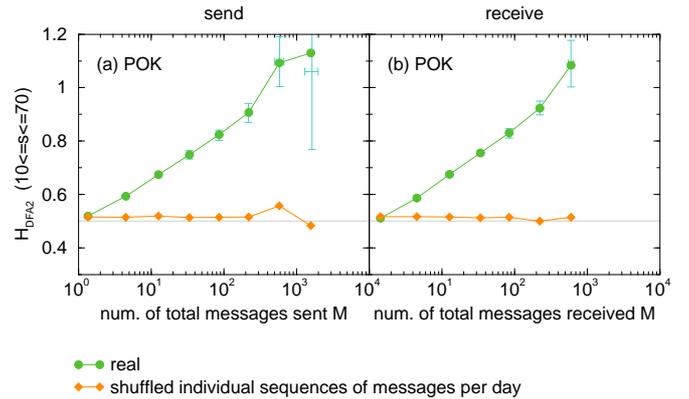}
\caption{
\label{fig:oc2dfaaveallbinfit2}
Fluctuation exponents of the communication activity (a)~sending and 
(b)~receiving messages by members of POK (weekly resolution). 
The exponents are plotted as a function of the activity level~$M$ for 
the original data (green circles), and 
individually shuffled sequences (orange diamonds).
See also~\cite{RybskiBHLM2009}.
}
\end{centering}
\end{figure}

First, we define the activity record, $\mu_j(t)$, counting 
the number of messages member~$j$ sends at day/week~$t$.
Thus, we study the activity that is aggregated at 
the daily or weekly level.
This is done to avoid possible oscillations that are observed in the data 
at both frequencies.

In a previous study \cite{RybskiBHLM2009} we have applied 
Detrended Fluctuation Analysis (DFA) 
\cite{PengBHSSG94,BundeHKPPV00,KantelhardtKRHB01} 
and found that the activity records, $\mu(t)$,
exhibit long-term correlations, which are characterized by a 
power-law decaying auto-correlation function, 
\begin{eqnarray}
C(\Delta t) & = & \frac{1}{\sigma_\mu^2}\left\langle
\left[\mu(t)-\langle\mu(t)\rangle\right]
\left[\mu(t+\Delta t)-\langle\mu(t)\rangle\right] 
\right\rangle \nonumber \\
& \sim & (\Delta t)^{-\nu} \nonumber
\, ,
\end{eqnarray}
where~$\langle\mu(t)\rangle$ is the average of the record~$\mu(t)$, 
$\sigma_\mu$ is its standard deviation, and 
$\nu$ is the correlation exponent ($1\ge\nu\ge0$).
The fluctuation function provided by DFA scales as 
\begin{equation}
\label{eq:FsimH}
F(\Delta t)\sim (\Delta t)^H
\end{equation}
where the exponent~$H$ is similar 
to the Hurst exponent 
($1/2\le H\le 1$, larger exponents correspond to 
more pronounced long-term correlations).
It is related to the correlation exponent via
\begin{equation}
\label{eq:h1g2}
\nu=2-2H
\, .
\end{equation}
For uncorrelated or short-term correlated records the asymptotic 
fluctuation exponent is $H=1/2$
(for a review we refer to \cite{KantelhardtJW2010}).

In order to study the activity with respect to long-term correlations, 
we apply second order DFA (DFA2) 
\cite{BundeHKPPV00,KantelhardtKRHB01} 
[linear detrending of $\mu_j(t)$] 
and obtain the fluctuation functions, $F^j_{\rm DFA2}(\Delta t)$ 
(details can be found in \cite{RybskiBHLM2009}).
Since the activity records of the individual members are too short, 
we average the squared fluctuation functions among members with 
similar overall activity (i.e. total number of messages, $M$): 
$F(\Delta t)=[\sum_{j|M}(F^j(\Delta t))^2]^{1/2}$. 
Therefore, we employ logarithmic bins in $M$. 
The activity distributions are discussed in Sec.~\ref{ssubsec:distributions}.

In Fig.~\ref{fig:dfaaveallbind1v2} we compare for sending in 
POK the fluctuation functions in 
daily resolution [Fig.~\ref{fig:dfaaveallbind1v2}(a)] and 
weekly resolution [Fig.~\ref{fig:dfaaveallbind1v2}(b)]. 
In order to match the scales, we have shifted the curves in
Fig.~\ref{fig:dfaaveallbind1v2}(b) along the $\Delta t$-axis. 
Naturally, in daily resolution, the fluctuation functions cover more scales. 
The asymptotic scaling is in both cases the same, 
namely no correlations in the case of 
least active members and strong long-term correlations with 
fluctuation exponents close to $1$ for the most active members. 
Moreover, for POK in daily resolution, 
the fluctuation functions exhibit an increase 
from small slopes on short time scales to larger slopes on large scales. 
This indicates that the long-term correlations do
not vanish after certain scale, but the opposite, the long-term
correlations become stronger. 
Note, that we use weekly resolution in order to cope with 
possible weekly oscillations \cite{GolderWH2006,LeskovecH2008,MalmgrenSMA2008}.

We measure the fluctuation exponents by applying least squares fits to 
$\log F(\Delta t)$ vs. $\log \Delta t$ on the scales 
$10<\Delta t<63$\,days (QX) and $10<\Delta t<70$\,weeks (POK). 
For the former case the obtained fluctuation exponents are 
plotted in Fig.~\ref{fig:oc1dfaaveallbinfit2} 
as a function of the members activity level, 
i.e. their total number of messages~$M$. 
For sending [panel~(a)], the less active members 
exhibit uncorrelated behavior. 
The more messages the members send overall, the stronger correlated
is their activity. 
The fluctuation exponent~$H_{\rm QX}$ increases 
with~$M$ and reaches values up to $0.75\pm 0.05$ (sending). 
In contrast, for the shuffled data, the fluctuation
exponents are always very close to $1/2$. 
This confirms that the long-term correlations are due to the 
temporal structure of the times each member sends his/her messages, 
see also \cite{RybskiBHLM5a}. 
For receiving messages, Fig.~\ref{fig:oc1dfaaveallbinfit2}(b), 
we find almost identical results.
The error bars in Fig.~\ref{fig:oc1dfaaveallbinfit2} were calculated by 
subdividing the groups of different activity level. 
The size of the error bars is simply the standard deviation of the
corresponding exponents.

The estimated fluctuation exponents obtained for POK are displayed in
Fig.~\ref{fig:oc2dfaaveallbinfit2}. 
Qualitatively, we obtain a similar picture as for QX. 
However, in contrast to QX, here the original records achieve larger 
fluctuation exponents up to $0.91 \pm 0.04$ 
(sending), disregarding the last points which carry large error-bars. 
A possible reason for these different maximum exponents 
could be that in the case of POK the data covers a much longer period of 
data acquisition, and possible non-stationarities \cite{VazquezA2007}.
In QX, the members might not have had enough time to exhibit the full 
extend of their persistence, while in POK we follow the entire 
evolution of the online community. 

Indeed, similar behavior of long-term correlations have been 
found in traded values of stocks and e-mail communication 
\cite{EislerK2006,EislerBK2008}, 
where the fluctuation exponent increases in an analogous way with 
the mean trading activity of the corresponding stock or with the 
average number of e-mails (see also \cite{RochaLH2010}).

Apart from these, for human related data, 
long-term persistence has been reported for 
physiological records \cite{PengMHHSG1993,IvanovBAHFBSG1999,BundeHKPPV00}, 
written language \cite{KosmidisKA2006}, 
or for records generated by collective behavior such as 
finance and economy \cite{LiuGCMPS1999,MantegnaS1999,LuxA2002}, 
Ethernet traffic \cite{LelandTWW1994}, 
Wikipedia access \cite{KaempfTKM2011}, 
as well as highway traffic \cite{TadakiKNNSSY2006,XiaoYanZHM2007}. 
There are also indications of long-term correlations in 
human brain activity \cite{LinkenkaerHansenNPI2001,AllegriniMBFGGWP2009} and 
human motor activity \cite{IvanovHHSS2007}.

A question that arises is, why the fluctuation exponent 
(in Figs.~\ref{fig:oc1dfaaveallbinfit2} and~\ref{fig:oc2dfaaveallbinfit2}) 
depends on the activity level of the members, that is, 
why the least active members exhibit no persistence 
while the most active members exhibit strong persistence. 
We argue that if only few messages appear in the whole period of 
data acquisition, long-term persistence cannot be reflected. 
In these cases it is quite possible that much longer records and higher
aggregation level such as months or years would be needed to reveal the
persistence. But doing so, there would be other members with even less messages
which then again would probably appear with seemingly uncorrelated 
message signals. 
Thus, we propose that the exponents of the largest activity reflect 
more accurately the scaling behavior of human communication activity. 
In Sec.~\ref{sssec:model} 
we propose statistical simulations to generate data using 
peaks over threshold (POT) and find that it supports 
this perception.

\begin{figure}
\begin{centering}
\includegraphics[width=\columnwidth]{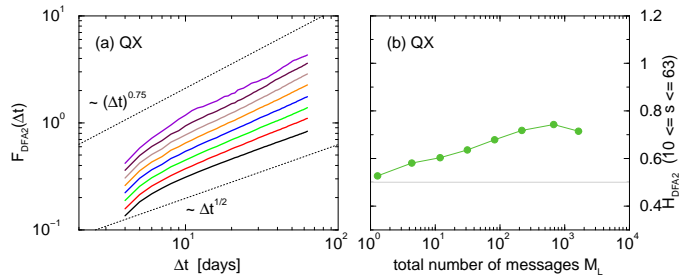}
\caption{
\label{fig:oc1dfalnk}
Temporal correlations in the daily amount of messages on \emph{directed links} 
in QX. 
(a)~DFA2 fluctuation functions versus the time scale $\Delta t$, averaged 
conditional to the final number of messages of each link.
The different curves correspond to different activity levels: 
$M_{\rm L}=$1-2, 3-7, 8-20, 21-54, 55-148, 149-403, 404-1096, 1097-2980 
(from bottom to top). 
The dotted lines correspond to the exponents $H=0.75$~(top) and 
$H=1/2$~(bottom).
(b)~The DFA2 fluctuation exponent $H_{\rm L, QX}$ obtained from~(a) 
is plotted as a function of the activity level~$M_{\rm L}$.
The exponents were obtained in the range of scales $10\le \Delta t\le63$\,days. 
The activity along directed links comprise similar long-term correlations 
as the total activity of individual members to all of their acquaintances.
}
\end{centering}
\end{figure}

At this point we need to mention that long-term correlations can be related 
to broad inter-event time distributions, 
i.e. the times between successive messages of individual members. 
Such distributions have been investigated, 
see e.g. \cite{BarabasiAL2005,MalmgrenSCA2009}, but there is no consensus 
on the functional form.
We study the inter-event time distributions in a different 
publication \cite{RybskiBHLM5a} where we demonstrate the 
connection with the long-term correlations found here.

\subsubsection*{Along directed links}

In Fig.~\ref{fig:oc1dfalnk} we study for QX the long-term correlations 
in activity not on the sender or receiver (node) level 
but on the level of messages along directed links. 
This means that we track when a message is sent directed 
between two members but 
separately for any pair of members, such as 
\texttt{a}$\rightarrow$\texttt{b}, 
\texttt{b}$\rightarrow$\texttt{a}, 
\texttt{a}$\rightarrow$\texttt{d}, \dots. 
Accordingly, we determine the activity records 
$\mu_{\texttt{ab}}(t)$, $\mu_{\texttt{ba}}(t)$, $\mu_{\texttt{ad}}(t)$, etc., 
expressing how many messages have been sent each day/week, $t$, between 
any pair of members.
Analogous, there is also an activity level for the links, 
$M^{\texttt{ab}}_{\rm L}$, \dots 
(we disregard those pairings without activity). 
Then we perform the analogous analysis for long-term correlations by 
applying DFA2 and averaging among pairings with similar 
overall activity 
(the distributions of activity are discussed in 
Sec.~\ref{ssubsec:distributions}).
The fluctuation functions in Fig.~\ref{fig:oc1dfalnk}(a) have
asymptotic slopes close to $1/2$ for those links with few total number of
messages. 
In contrary, those links with many total number of messages exhibit
long-term correlations with exponents up to~$0.74$. 
The fluctuation exponents as a function of the activity level~$M_{\rm L}$ 
are plotted in Fig.~\ref{fig:oc1dfalnk}(b).
Apart from the fact, that by definition the number of messages on the most
active links is lower (or equal) than the number of messages of the most active 
members, the curve looks very similar to the one in 
Fig.~\ref{fig:oc1dfaaveallbinfit2}(a), 
in particular the maximum exponents are quite similar 
($H_{\rm QX}\approx 0.75$ and $H_{\rm L, QX}\approx 0.74$).
This indicates, that the persistence in the communication 
may be dominated by the communication activity with the most liked 
partners.

In \cite{HidalgoR2008} a different concept of persistence links has been 
investigated. The period of data acquisition is partitioned into time slices 
in each of which a network is built. Then the persistence is defined as the 
normalized number of time slices in which a certain link appears. 
However, the approach of our work is not compatible with the one in 
\cite{HidalgoR2008} and we cannot directly compare the results.

\subsection{Growth process}
\label{subsec:resgrw}

\subsubsection{Growth in the number of messages}
\label{ssubsec:resgrwmes}

As suggested in \cite{RybskiBHLM2009} we also analyze the growth properties of 
the message activity. 
This concept is borrowed from econophysics, where the growth of companies 
has been found to exhibit non-trivial scaling laws \cite{StanleyABHLMSS1996}, 
that in particular violate the original Gibrat's law 
\cite{GibratR1931,SuttonJ1997,MitzenmacherM2004,SaichevMS2009} 
and at the same time represents a generalized Gibrat's law (GGL) 
\cite{RybskiBHLM2009}. 
In the present study, each member is considered as a unit and the number of 
messages sent or received since the beginning of data acquisition represents 
its size.
We analyze the growth in the number of messages in analogy to other 
systems such as the growth of companies 
\cite{StanleyABHLMSS1996,AmaralBHSS1998} or 
the growth of cities \cite{RozenfeldRABSM2008,RozenfeldRGM2009}. 
The analogy is supported by some aspects:
(i) The members of a community represent a population similar to the 
population of a country.
(ii) The number of members fluctuates and typically grows analogous to the 
number of cities of a country.
(iii) The activity or number of links of individuals fluctuates and 
grows similar to the size of cities.

The cumulative number, $m^j(t)$, expresses how many messages have been
sent by a certain member~$j$ up to a given time~$t$ [for a better
readability we will not write the index~$j$ explicitly, $m(t)$]. 
We consider the evolution of $m(t)$ between times~$t_0$ and~$t_1$ 
within the period of data acquisition~$T$ ($t_0<t_1\le T$)
as a growth process, where each member exhibits a
specific growth rate~$r_j$ ($r$~for short notation):
\begin{equation}
r=\ln\frac{m_1}{m_0}
\, ,
\label{eq:rlnm1m0}
\end{equation}
where $m_0\equiv m(t_0)$ and $m_1\equiv m(t_1)$ are the number of
messages sent until $t_0$ and $t_1$, respectively, by every member.
To characterize the dynamics of the activity, we consider two
measures. 
(i) The conditional average growth rate, $\langle r(m_0)\rangle$, 
quantifies the average growth of the number of messages sent by the 
members between~$t_0$ and~$t_1$ depending on the initial number of 
messages, $m_0$. 
In other words, we consider the average growth rate of only those members 
that have sent~$m_0$ messages until~$t_0$. 
(ii) The conditional standard deviation of the growth rate for those members 
that have sent~$m_0$ messages until~$t_0$, 
\begin{equation}
\sigma(m_0)\equiv\sqrt{\langle(r(m_0)-\langle r(m_0)\rangle)^2\rangle}
\, , 
\end{equation}
expresses the statistical spread or fluctuation of growth among the members 
depending on $m_0$. 
Both quantities are relevant in the context of Gibrat's law in economics
\cite{GibratR1931,SuttonJ1997,MitzenmacherM2004,SaichevMS2009}
which proposes a proportionate growth process entailing the assumption
that the average and the standard deviation of the growth rate of a
given economic indicator are constant and independent of the specific
indicator value.
That is, both $\langle r(m_0)\rangle$ and $\sigma(m_0)$ are independent of
$m_0$.

As shown in \cite{RybskiBHLM2009}, for the message data 
the conditional average growth rate is 
almost constant and only decreases slightly, 
\begin{equation}
\langle r(m_0)\rangle \sim m_0^{-\alpha}
\, ,
\label{eq:rmosimm0a}
\end{equation}
with an exponent $\alpha\approx 0.05$. 
This means that members with many messages in average
increase their number of messages almost with the same rate as members with few
messages. In contrast, the conditional standard deviation clearly decreases with
increasing $m_0$, 
\begin{equation}
\sigma(m_0) \sim m_0^{-\beta}
\label{eq:sm0simmob}
\, ,
\end{equation}
where $t_0=T/2$ is optimal in terms statistics. 
In this case the exponents $\beta_{\rm QX}=0.22\pm 0.01$ and 
$\beta_{\rm POK}=0.17\pm 0.03$ for sending messages 
were found \cite{RybskiBHLM2009}. 
This means, although the average growth rate almost does not depend 
on $m_0$, the conditional standard deviation of the growth of members
with many messages is smaller than the one of members with few messages.
Due to weaker fluctuations, active members are relatively 
better predictable in their activity of sending messages.

It has been shown that the fluctuation exponent~$H$ and the 
growth fluctuation exponent~$\beta$ are related via
\cite{RybskiBHLM2009}
\begin{equation}
\beta=1-H
\, .
\label{eq:beta1h}
\end{equation}
Equation~(\ref{eq:beta1h}) is a scaling law formalizing the relation 
between growth and long-term correlations in the activity. 
According to Eq.~(\ref{eq:beta1h}), the original Gibrat's law
($\beta_{\rm G}=0$) corresponds to very strong long-term correlations
with $H_{\rm G}=1$. 
In contrast, $\beta_{\rm rnd}=1/2$ represents completely random activity 
$(H_{\rm rnd}=1/2)$.
The observed message data comprises $1/2>\beta>0$ and $1/2<H<1$.
Surprisingly, the values of $\beta$ found here are very close to 
the $\beta$ values found for companies 
in the US economy \cite{StanleyABHLMSS1996}. 

In the case of companies, also the distribution of growth rates 
has been studied. 
It was found that the distribution density 
follows \cite{StanleyABHLMSS1996}:
\begin{equation}
p(r|m_0)=\frac{1}{s\sigma(m_0)}
\exp\!\left(-\frac{s|r-\langle r(m_0)\rangle|}{\sigma(m_0)}\right)
\, ,
\label{eq:prtent}
\end{equation}
whereas $s=\sqrt{2}$.
Next we analyze, how the growth rates~$r$ are distributed in the case of 
the message data. 

First we need to point out that in contrast to the growth of companies, 
our entities can never shrink. 
The members cannot loose messages, 
the number~$m(t)$ either increases or remains the same. 
Accordingly, in our case $r\ge 0$ and therefore $s=1$, as can be 
derived for the single-sided exponentially decaying distribution.

\begin{figure}
\begin{centering}
\includegraphics[width=\columnwidth]{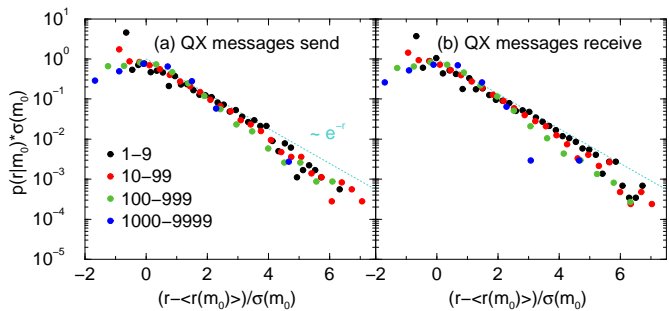}
\caption{
\label{fig:oc1mestnt}
Scaled probability density of growth rates $r$, Eq.~(\ref{eq:rlnm1m0}), 
in the number of messages by members of QX.
(a)~Sending and (b)~receiving.
The times for $m_0$ and $m_1$ have been chosen as $t_0=T/2$ and $t_1=T$.
The symbols correspond to different initial number of messages $m_0$.
The axis are scaled assuming a distribution according to 
Eq.~(\ref{eq:prtent}) with $s=1$ which then corresponds to the dotted lines.
}
\end{centering}
\end{figure}

Figure~\ref{fig:oc1mestnt} shows $p(r|m_0)$ for QX where the values are 
scaled to collapse according to Eq.~(\ref{eq:prtent}) with $s=1$. 
In order to have reasonable statistics, 
we define the condition~$m_0$ in rather wide ranges, 
namely according to the decimal logarithm. 
For sending [Fig.~\ref{fig:oc1mestnt}(a)] and 
receiving [Fig.~\ref{fig:oc1mestnt}(b)] messages 
the scaled probability densities collapse and are quite similar. 
Nevertheless, the growth rates do not exactly follow Eq.~(\ref{eq:prtent}) 
with $s=1$. 
While for the less active members with small growth rates we find a good 
agreement, for more active members and large growth rates the obtained curves 
deviate from the theoretical one towards a steeper decay.

\begin{figure}
\begin{centering}
\includegraphics[width=\columnwidth]{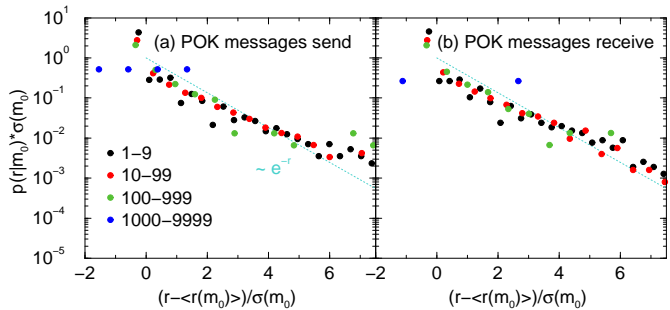}
\caption{
\label{fig:oc2mestnt}
Scaled probability density of growth rates $r$, Eq.~(\ref{eq:rlnm1m0}), 
in the number of messages by members of POK.
(a)~Sending and (b)~receiving.
Analogous to Fig.~\ref{fig:oc1mestnt}.
}
\end{centering}
\end{figure}

The corresponding results for POK are shown in Fig.~\ref{fig:oc2mestnt}.
Again, sending and receiving are very similar. 
The curves collapse reasonably, but in contrast to QX here the 
measured $p(r|m_0)$ overall deviate from the theoretical one comprising 
less steep slopes.

We argue that as for single time series, distribution and correlation 
properties are in most cases independent, the same holds for the 
message data and the growth.
The distribution of growth rates~$p(r|m_0)$ seems to be independent 
from the long-term correlations which are reflected in $\sigma(m_0)$ 
with the exponent~$\beta$.

\subsubsection{Mutual growth in the number of messages}
\label{ssubsec:resmutgrwmes}

Next we study a variation of growth. 
Instead of considering the absolute number of messages a member sends, 
we study the difference in the number of messages compared to any other member, 
the mutual difference $m^i(t)-m^j(t)$.
Thus, the growth rate is defined analogous to Eq.~(\ref{eq:rlnm1m0})
\begin{equation}
r_\times=\ln\frac{m_1^i-m_1^j}{m_0^i-m_0^j}
\label{eq:rxm1im1j}
\end{equation}
where now there is a growth rate for every pair of members~$i$ and~$j$. 
The conditional average growth rate and the corresponding 
standard deviation is then taken over all possible pairs and the 
condition is the difference at~$t_0$,
$m^i_0-m^j_0=m^i(t_0)-m^j(t_0)$, providing the quantities 
$\langle r_\times(m^i_0-m^j_0)\rangle$ and $\sigma(m^i_0-m^j_0)$.
We disregard combinations of~$i$ and~$j$ where 
$m_0^i-m_0^j=0$ or 
$\frac{m_1^i-m_1^j}{m_0^i-m_0^j}\le0$.

\begin{figure}
\begin{centering}
\includegraphics[width=\columnwidth]{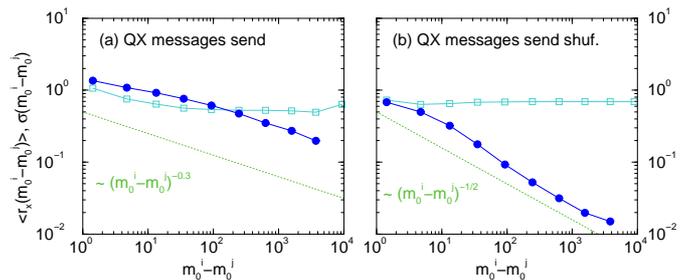}
\caption{
\label{fig:xmibrat}
Average mutual growth rate and standard deviation versus foregoing difference 
in the number of messages for sending in QX.
The average (open squares) and standard deviation (filled circles) 
of the mutual growth rate $r_\times$, Eq.~(\ref{eq:rxm1im1j}), are plotted 
conditional to the initial difference $m_0^i-m_0^j$, whereas $t_0=T/2$ and 
$t_1=T$.
(a)~Original data and (b)~shuffled data.
The dotted line in~(a) corresponds to the exponent $\beta_\times=0.3$ and 
in~(b) to $\beta_\times=1/2$.
}
\end{centering}
\end{figure}

The results for sending in QX are shown in Fig.~\ref{fig:xmibrat}. 
Apart from a small decrease up to $m^i_0-m^j_0\approx 50$, 
the average growth rate is constant [Fig.~\ref{fig:xmibrat}(a)]. 
The conditional standard deviation asymptotically follows a 
slope~$\beta_\times\approx 0.3$ with deviations to small exponents for 
small~$m^i_0-m^j_0$. 
In the case of the shuffled data [Fig.~\ref{fig:xmibrat}(b)], 
as expected, the average growth rate is constant while the standard 
deviation decreases steeper than for the original data, namely with 
$\beta_{\times,{\rm rnd}}\simeq 1/2$, although not with a nice straight line. 
Nevertheless, we conclude that the scaling of the standard
deviation in Fig.~\ref{fig:xmibrat}(a) must be due to temporal correlations 
between the members. The growth of the difference between their number of 
messages comprises similar scaling as the individual growth.

We conjecture that $\sigma(m^i_0-m^j_0)$ reflects long-term cross-correlations 
in analogy to $\sigma(m_0)$ for auto-correlations.
However, so far, we are not able to provide further evidence 
for this analogy and the corresponding relation to 
$\beta=1-H$, Eq.~(\ref{eq:beta1h}), 
since an appropriate technique for the direct quantification of 
long-term cross-correlations is lacking.

\subsection{Modeling}
\label{subsec:modeling}

In what follows, we propose numerical simulations with the purpose 
of testing the methods and empirical patterns we found.
We study three approaches adopted to the modeling of human activity:
(a) peaks over thresholds,
(b) preferential attachment \cite{BarabasiA1999}, and 
(c) cascading Poisson process \cite{MalmgrenSMA2008}.

\subsubsection{Peaks over threshold (POT) simulations}
\label{sssec:model}

Our finding that the activity of sending messages exhibits 
long-term persistence asserts the existence of an 
underlying long-term correlated process. 
This can be understood as an unknown individual state driven by 
various internal and external stimuli 
\cite{HedstroemP2005,KentsisA2006,PallaBV2007,CraneS2008,MalmgrenSMA2008,AllegriniMBFGGWP2009} 
increasing the probability to send messages. 
Generating such a hypothetical long-term correlated internal process~$(x_i)$, 
simulated message data can be defined by the instants at which 
this internal process exceeds a threshold~$q$ (peaks over threshold, POT), 
see \cite{BundeEKH2005,AltmannK2005,EichnerKBH2007} and references therein.

More precisely, we consider a long-term correlated 
sequence~$(x_i)$ consisting of $N^*$~random numbers 
that is normalized to zero average ($\langle x\rangle=0$) and 
unit standard deviation ($\sigma_x=1$). 
Choosing a threshold~$q$, at each instant~$i$ the probability to 
send a messages is:
\begin{equation}
p_{\rm snd} = 
\left\{\begin{array}{cl} 
1 & \mbox{ for }x_i>q\\ 
0 & \mbox{ for }x_i\le q 
\end{array}\right.
\, .
\end{equation}
Thus, the message events are given by the indices~$i$ 
of those random numbers~$x_i$ exceeding~$q$.

\begin{figure}
\begin{centering}
\includegraphics[width=0.7\columnwidth]{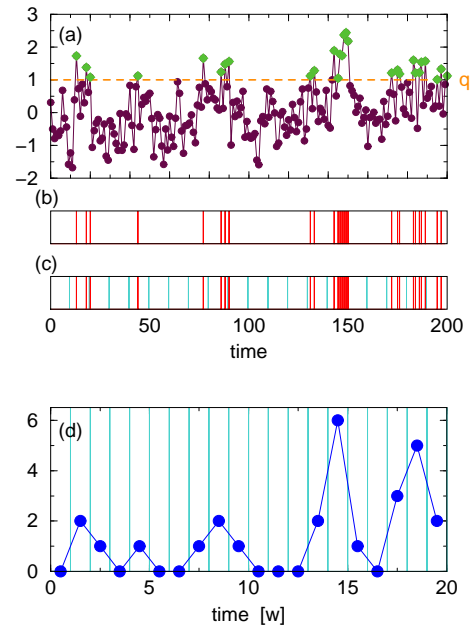}
\caption{
\label{fig:millu}
Illustration of the peaks over threshold simulations. 
(a)~An underlying and unknown long-term correlated process 
determines the instantaneous probability of sending messages.
Once this state passes certain threshold~$q$ (dashed orange line) 
a messages is sent (green diamonds).
(b)~Generated instants of messages, 
(c)~with windows for aggregation, such as messages per day.
(d)~Aggregated record of messages in windows of size~$w$, 
here $w=10$.
}
\end{centering}
\end{figure}

Figure~\ref{fig:millu}(a) illustrates the procedure. 
The random numbers are plotted as brown circles and the events 
exceeding the threshold (orange dashed line) by the green diamonds.
The resulting instants are depicted in Fig.~\ref{fig:millu}(b) 
representing the simulated messages. 
The threshold approximately predefines the total number of events 
and accordingly the average inter-event time. 
Using normal-distributed numbers~$(x_i)$, the number of events/messages 
is approximately given by the length~$N^*$ and 
the inverse cumulative distribution function associated with 
the standard normal distribution (probit-function). 
Additionally, the random numbers we use are long-term correlated 
with variable fluctuation exponent. 
We impose these auto-correlations using Fourier Filtering Method 
\cite{KantelhardtKRHB01,MakseHSS96}. 
Next we show that this process reproduces the scaling 
in the growth, i.e. GGL, as well as the variable long-term correlations in 
the activity of the members 
(e.g. Figs.~\ref{fig:oc1dfaaveallbinfit2} and~\ref{fig:oc2dfaaveallbinfit2}).

\begin{figure}
\begin{centering}
\includegraphics[width=\columnwidth]{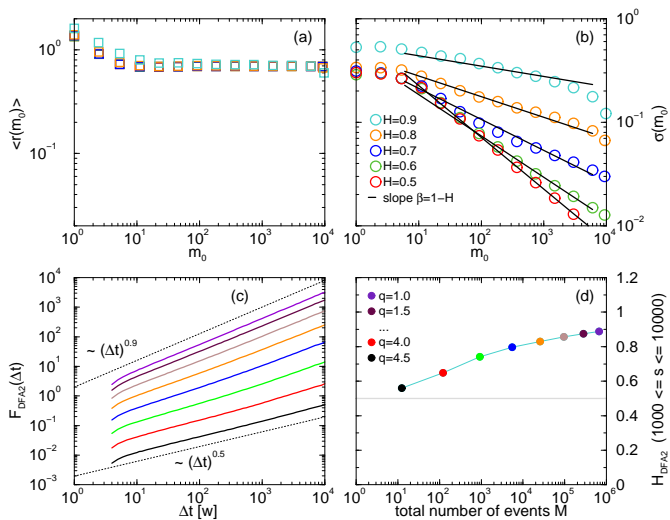}
\caption{
\label{fig:model}
Results of numerical simulations.
(a)~Mean growth rate conditional to the number of events until~$t_0=N^*/2$ 
as obtained from~$100,000$ long-term correlated records of length~$N^*=131,072$ 
with variable imposed fluctuation exponent~$H_{\rm imp}$ 
between~$1/2$ and~$0.9$ and random threshold~$q$ between~$1.0$ and~$6.0$.
(b)~As before but standard deviation conditional to the number of events. 
The solid lines represent power-laws with exponents $\beta$ expected from the 
imposed long-term correlations according to Eq.~(\ref{eq:beta1h}).
(c)~Long-term correlations in the sequences of aggregated peaks 
over threshold. 
For every threshold~$q$ between~$1.0$ (violet) and~$4.5$ (black) 
$100$~normalized records of length~$N^*=4,194,304$ have been created with 
$H_{\rm imp}=0.9$. 
The events are aggregated in windows of size $w=100$. 
The panel shows the averaged DFA2 fluctuation functions. 
(d)~Fluctuation exponents on the scales $1,000\le s \le 10,000$, 
as a function of the total number of events.
}
\end{centering}
\end{figure}

For testing this process we create 100,000 independent long-term 
correlated records~$(x_i)$ of length $N^*=131,072$, 
impose the fluctuation exponent~$H_{\rm imp}$, 
and choose for each one a random threshold~$q$ between~$1$ and~$6$, 
each representing a sender. 
Extracting the peaks over threshold, we obtain the events and determine 
for each record/member the growth in the number of events/messages 
between~$N^*/2$ and~$N^*$. 
This is, for each record/member we count the numbers of events/messages 
$m_0$ until $t_0=t_{i=N^*/2}$ as well as $m_1$ until $t_1=t_{i=N^*}$ and 
calculate the growth rate according to Eq.~(\ref{eq:rlnm1m0}). 
We then calculate the conditional average $\langle r(m_0)\rangle$ and the 
conditional standard deviation $\sigma(m_0)$ where the values of $m_0$ are 
binned logarithmically. 
The quantities are plotted in Fig.~\ref{fig:model}(a) and~(b), 
while in panel~(b) we include slopes expected from $\beta=1-H$, 
Eq.~(\ref{eq:beta1h}). 
We find that the numerical results reasonably agree with the 
prediction (solid lines). 
Except for small $m_0$, these results are consistent 
with those found in the original message data.

The fluctuation functions can be studied in the same way. 
As described in Sec.~\ref{subsec:resltc}, 
we find long-term correlations in the sequences of 
messages per day or per week. 
On the basis of the above explained simulated messages, 
we analyze them in an analogous way. 
For each threshold $q=1.0, 1.5, \dots 4.0, 4.5$ we 
create $100$~long-term correlated records of length $N^*=4,194,304$ with 
imposed fluctuation exponent~$H_{\rm imp}=0.9$, 
extract the simulated message events, 
and aggregate them in non-overlapping windows of size $w=100$. 
This is, tiling $N^*$ in segments of size~$w$ and counting the number of 
events occurring in each segment [Fig.~\ref{fig:millu}(c) and~(d)]. 
The obtained aggregated records represent the analogous of messages per day or 
per week and are analyzed with DFA averaging the fluctuation functions 
among those configurations with the same threshold and thus similar 
number of total events.
The corresponding results are shown in Fig.~\ref{fig:model}(c) and~(d). 
We obtain very similar results as in the original data. 
We find vanishing correlations 
for the sequences with few events (large~$q$) and pronounced 
long-term correlations for the cases of many events (small~$q$), 
while the maximum fluctuation exponent corresponds to the 
chosen~$H_{\rm imp}$.
This can be understood by the fact that for $q$~close to zero the sequence 
of number of events per window converges to the aggregated sequence 
of~$0$ or~$1$ (for $x\le 0$ or $x>0$) reflecting the same 
long-term correlation properties as the 
original record \cite{XuYMSBI2010}. 
For a large threshold~$q$ too few events occur to measure the correct long-term 
correlations, e.g. the true scaling only turns out on larger unaccessible 
time scales requiring larger $w$ and longer records.

Although the simulations do not reveal the origin of 
the long-term correlated patchy behavior, 
they support Eq.~(\ref{eq:beta1h}) and the concept of an 
underlying long-term correlated process. 
Consistently, an uncorrelated, completely random, underlying process 
recovers Poisson statistics and therefore 
$\beta_{\rm rnd}=1/2$ for the growth fluctuations 
as well as uncorrelated message activity ($H_{\rm rnd}=1/2$).
For $1/2<H<1$ it has been shown \cite{BundeEKH2005,EichnerKBH2007} 
that the inter-event times follow a stretched exponential, 
see also \cite{BarabasiAL2005,StehleBB2010,RybskiBHLM5a}.

\subsubsection{Preferential attachment}
\label{sssubsec:bamodel}

Next we compare our findings with the growth properties of 
a network model. 
We investigate the Barabasi-Albert (BA) model which is based 
on preferential attachment and has been introduced to generate 
a kind of scale-free networks \cite{BarabasiA1999,AlbertB2002} with 
power-law degree distribution~$p(k)$ \cite{EbelMB2002,NewmanFB2002}, 
whereas the degree~$k$ of a node is the number of links it has to 
other ones.
Essentially, it consists of subsequently adding nodes
to the network by linking them to existing nodes which are chosen
randomly with a probability proportional to their degree.

We obtain the undirected network and study the degree growth properties by 
calculating the conditional average growth rate 
$\langle r_{\rm BA}(k_0)\rangle$ 
and the conditional standard deviation $\sigma_{\rm BA}(k_0)$ 
obtained from the scale-free BA model. 
The times~$t_0$ and~$t_1$ are defined by the number of nodes attached to
the network.

\begin{figure}
\begin{centering}
\includegraphics[width=0.8\columnwidth]{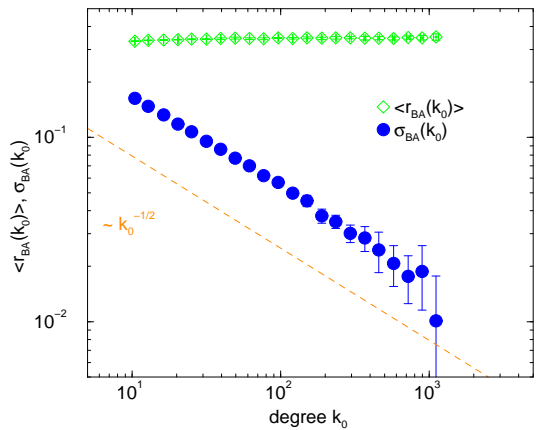}
\caption{
\label{fig:bamodel}
Average degree growth rate and standard deviation versus foregoing degree 
for the preferential attachment network model \cite{BarabasiA1999}.
The average (green open diamonds) and standard deviation (blue filled circles) 
of the growth rate~$r_{\rm BA}$ are plotted conditional to $k_0$, 
the degree of the corresponding nodes at the first stage.
We choose average degree $\langle k\rangle =20$, $50,000$~nodes in~$t_0$, 
and $100,000$~nodes in~$t_1$.
The error-bars are taken from $10$ configurations.
The dashed line in the bottom corresponds to~$\beta_{\rm BA}=1/2$.
}
\end{centering}
\end{figure}

Figure~\ref{fig:bamodel} shows the results where 
an average degree $\langle k\rangle =20$; 
$50,000$~nodes in~$t_0$, and $100,000$ nodes in~$t_1$ 
were chosen.
We find constant average growth rate that does not depend on the 
initial degree $k_0$.
The conditional standard deviation is a function of $k_0$ 
and exhibits a power-law decay with $\beta_{\rm BA}=1/2$ 
as expected for such an uncorrelated growth process 
\cite{RybskiBHLM2009}.
Therefore, a purely preferential attachment type of growth is not 
sufficient to describe the type of social network dynamics found in 
Sec.~\ref{ssubsec:resgrwmes}, since additional temporal correlations are 
involved in the dynamics of establishing acquaintances in the community.

The value $\beta_{\rm BA}=1/2$ in Eq.~(\ref{eq:beta1h}) corresponds 
to $H=1/2$ indicating complete randomness. 
There is no memory in the system. 
Since each addition of a new node is completely independent from 
precedent ones, 
there cannot be temporal correlations in the activity of adding links. 
In contrast, for the out-degree in QX and POK we obtained 
$\beta_{k,{\rm QX}}=0.22\pm0.02$ and 
$\beta_{k,{\rm POK}}=0.17\pm0.08$ 
\cite{RybskiBHLM2009}, which is supported by the 
(non-linear) correlations between the number of messages and the out-degree as 
presented in Sec.~\ref{subsec:resoth}.

Interestingly, an extension of the standard BA model has been proposed 
\cite{BianconiB2001}, see also \cite{BlasioSL2007,CohenH2009}, 
that takes into account different fitnesses of the nodes to acquiring links. 
We think that such fitness could be related to growth fluctuations, 
thus providing a route to modify the BA model to include the 
long-term correlated dynamics found here.

\subsubsection{Cascading Poisson process}
\label{ssubsec:cascpoisproc}

In this Section we elaborate the model proposed in \cite{MalmgrenSMA2008} and 
examine it with respect to long-term correlations.
The model is based on a cascading Poisson process (CPP), 
according to which the probability that a member enters an active interval is 
$\rho(t) = N_{\rm w} p_{\rm d}(t) p_{\rm w}(t)$, where 
$N_{\rm w}$ is the average number of active intervals per week, 
$p_{\rm d}(t)$ is the probability of starting an active interval at a 
particular time of the day, and 
$p_{\rm w}(t)$ is the probability of starting an active interval at a 
particular day of the week.
Once a member enters such an active interval he/she sends a set of 
$N_{\rm a}+1$ messages, where $N_{\rm a}$ is drawn from the distribution 
$p(N_{\rm a})$. The messages sent in such an active interval are sent 
randomly, i.e. a homogeneous Poisson process with rate $\rho_{\rm a}$ 
events per hour.

\begin{figure*}
\begin{centering}
\includegraphics[width=0.7\textwidth]{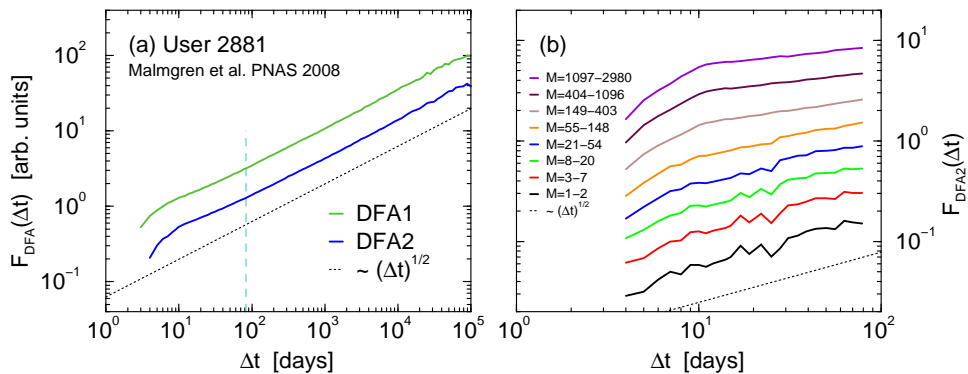}
\caption{
\label{fig:mmodel}
DFA fluctuation functions of message data created with the model proposed in 
\cite{MalmgrenSMA2008}. 
(a) User 2881. We visually extract the model parameters from 
Fig.~3 in \cite{MalmgrenSMA2008} and generate message data for 
approx. $800$k\,days. The panel shows the fluctuation functions from 
DFA1 and DFA2, which asymptotically go as $\sim(\Delta t)^{1/2}$, 
i.e. no long-term correlations. The hump on small scales is due to 
the model inherent oscillations \cite{KantelhardtKRHB01}. 
The dashed vertical line is placed at $\Delta t=83$\,days.
(b) Random parameterization. We randomly choose the model parameters, 
create $20$k records of $83$\,days and average the obtained 
DFA2 fluctuation functions according to the final number of messages, $M$.
While for those simulated members with few messages we find 
$F(\Delta t)\sim(\Delta t)^{1/2}$, the $F(\Delta t)$ of the most active 
members exhibit a hump due to oscillations similar to the one in panel (a).
}
\end{centering}
\end{figure*}

First, we study the example of User 2881 as analyzed in \cite{MalmgrenSMA2008}
(please note that in \cite{MalmgrenSMA2008} a different data set is studied and 
the user is neither in OC1 nor in OC2). 
We extract from \cite{MalmgrenSMA2008} the parameters 
$N_{\rm w}=7.3$ active intervals per week, 
$\rho_{\rm a}=1.7$ events per hour, as well as (visually) the 
distributions $p_{\rm d}(t)$, $p_{\rm w}(t)$, and $p(N_{\rm a})$.
The original period of 83 days is not sufficient to apply DFA and 
we run the model for this set of parameters over $800$k\,days. 
Then we extract the record of number of messages per day, $\mu(t)$, and 
apply DFA.
The obtained fluctuation functions are shown in Fig.~\ref{fig:mmodel}(a). 
On small scales below $100$\,days a hump in the $F(\Delta t)$ can be 
identified, which is due to oscillations in $\mu(t)$ \cite{KantelhardtKRHB01}. 
While asymptotically the influence of oscillations vanishes, 
on scales below the wavelength, the oscillations appear as correlations 
[increased slope in $F(\Delta t)$] and 
on scales above the wavelength, the oscillations appear as anti-correlations 
[decreased slope in $F(\Delta t)$].
Asymptotically, we find $F(\Delta t)\sim(\Delta t)^{1/2}$, 
i.e. $H_{\rm CPP}\simeq 1/2$, 
corresponding to a lack of long-term correlations.
Even on scales up to $83$\,days, we rather find $H_{\rm CPP}<1/2$ 
(which we expect from the imposed weekly oscillations).

Next, we study 20,000 simulated e-mail senders with randomly chosen parameters. 
(i) We fill $p_{\rm w}(t)$ with random numbers and set $p_{\rm w}(t)=0$ for 
$t=0,6$, i.e. Sunday and Saturday.
(ii) We fill $p_{\rm d}(t)$ with random numbers and set $p_{\rm d}(t)=0$ for 
$t=0\dots5$ and $t=23$, i.e. at night.
(iii) We set $p(N_{\rm a})$ starting with a random $p(N_{\rm a}=0)$. 
Then $p(N_{\rm a})$ decays exponentially up to a random $N_{\rm a}$ 
below $36$. 
$p_{\rm w}(t)$, $p_{\rm w}(t)=0$, and $p(N_{\rm a})$ are normalized. 
(iv) We randomly choose $0<N_{\rm w}<40$.
(v) We randomly choose $0<\rho_{\rm a}<30$. 
From \cite{MalmgrenSMA2008} Supporting Information 2 (SI2) we 
estimated the typical maximum values of $N_{\rm a}$, $N_{\rm w}$, 
and $\rho_{\rm a}$ ($36$, $40$, and $30$, respectively).
We run the model for $83$\,days and extract the $\mu(t)$ for each 
simulated e-mail sender.
Then we apply DFA2 and average the fluctuation functions according to the 
final number of messages, $M$, see also \cite{RybskiBHLM2009}.
The fluctuation functions for the various activity levels are depicted 
in Fig.~\ref{fig:mmodel}(b). 
We find that members with small final number of messages exhibit uncorrelated 
behavior. The more active the members the more pronounced become the 
oscillations which we already discussed in the context of 
Fig.~\ref{fig:mmodel}(a). 
Thus, asymptotic $H_{\rm CPP}<1/2$ for large $M$ is due to the 
weekly cycles \cite{KantelhardtKRHB01}.
In our data, oscillations do not dominate the DFA fluctuation functions. 
Moreover, for OC2 we also find long-term correlations in 
weekly resolution (Fig.~\ref{fig:dfaaveallbind1v2}).

Based on periodic probabilities and Poisson statistics, the CPP model 
represents a powerful concept to characterize inter-event times. 
For this purpose, the average $N_{\rm w}$ seems to be sufficient. 
However, in order to recover long-term correlations, 
time dependent $N_{\rm w}=N_{\rm w}(t)$ seem to be necessary. 
In fact, the number of active intervals per week, $N_{\rm w}$, 
fluctuates, as can be seen in \cite{MalmgrenSMA2008}SI2 
(upper most row of the panels).
Thus, we suggest to extend the model by introducing a memory kernel, 
see e.g. \cite{CraneS2008}, or by using long-term correlated $N_{\rm w}(t)$.

\subsection{Other correlations}
\label{subsec:resoth}

\begin{figure}
\begin{centering}
\includegraphics[width=\columnwidth]{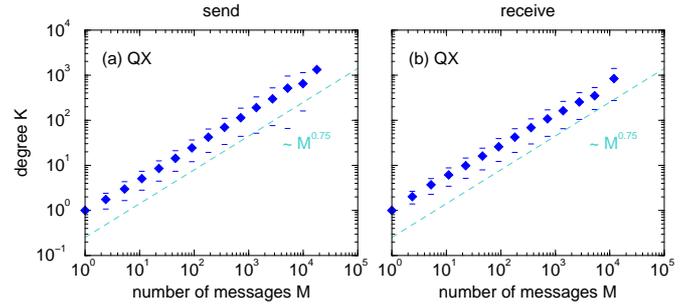}
\caption{
\label{fig:ccmibrat1}
Correlations between the final degree~$K$ and the final number of messages~$M$ 
for QX.
(a)~Out-degree and sending messages; 
(b)~in-degree and receiving messages.
The dashed lines correspond to a power-law with exponent $0.75$.
Members sending many messages also tend to have high out-degree, 
but not linearly, rather following a power-law.
}
\end{centering}
\end{figure}

In this Section we want to discuss other types of correlations. 
Figure~\ref{fig:ccmibrat1} shows for QX the final degree $K=k(T)$ versus the 
final number of messages $M=m(T)$. 
We find that for both, sending and receiving, 
the two quantities are correlated 
according to:
\begin{equation}
\label{eq:lambda}
K\sim M^\lambda \quad\rm{with}\quad \lambda\approx 3/4
\end{equation}
Similar relations have also been found for other
data \cite{MuchnikMH2010}. 
Since the correlations are positive, 
those members that send many messages, in average, 
also have more acquaintances to whom they send, but they know 
less acquaintances than they would in the case of linear 
correlations.
For receiving, Fig.~\ref{fig:ccmibrat1}(b), this correlation is very 
similar.

\begin{figure}
\begin{centering}
\includegraphics[width=\columnwidth]{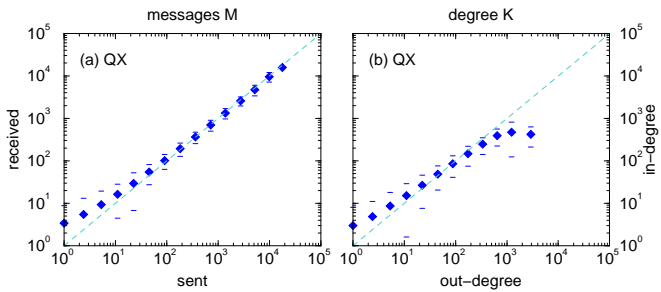}
\caption{
\label{fig:ccmibrat5}
Correlations between activity and passivity for QX.
(a)~Final number of messages $M$ received and sent;
(b)~final in- and out-degree.
The dashed lines correspond to a linear relation.
Those members who send many messages also receive many.
But, those members who know many people to whom they send messages do not 
necessarily know as many people from whom they receive messages.
}
\end{centering}
\end{figure}

The number of messages sent versus the number of messages received (for QX) 
is displayed in Fig.~\ref{fig:ccmibrat5}(a). Asymptotically the activity and 
passivity are linearly related and on average for every message sent 
there is a received one or vice versa. 
This, of course, does not mean that every message is replied. 
However, the less active members in average tend to receive more messages 
than they send. For example, those members who send in average one message 
receive about three. 
Nevertheless, the more active the members are the more the 
sending and receiving behavior approaches the linear relation.
In contrast, for the degree, Fig.~\ref{fig:ccmibrat5}(b), 
the asymptotic linear relation does not hold. 
Those members with large out-degree and small in-degree are referred to 
as spammers, since they send to many different people but only receive from 
few.

\begin{figure}
\begin{centering}
\includegraphics[width=\columnwidth]{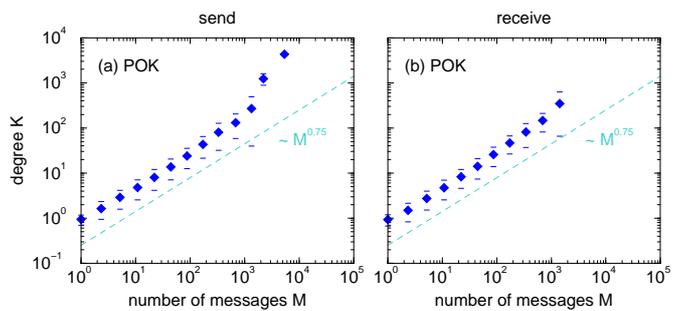}
\caption{
\label{fig:ccpibrat1}
Correlations between the final degree~$K$ and the final number of messages~$M$ 
for POK.
(a)~Out-degree and sending messages; 
(b)~in-degree and receiving messages.
Analogous to Fig.~\ref{fig:ccmibrat1}.
}
\end{centering}
\end{figure}
\begin{figure}
\begin{centering}
\includegraphics[width=\columnwidth]{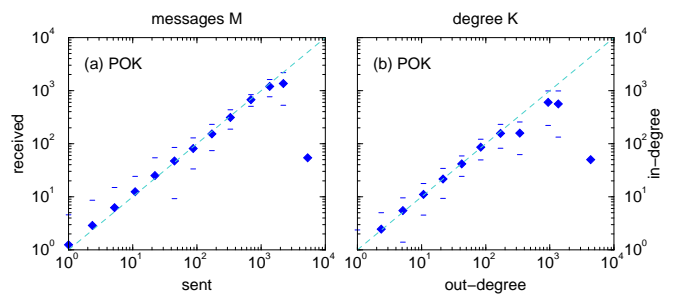}
\caption{
\label{fig:ccpibrat5}
Correlations between activity and passivity for POK.
(a)~Final number of messages $M$ received and sent;
(b)~final in- and out-degree.
Analogous to Fig.~\ref{fig:ccmibrat5}
}
\end{centering}
\end{figure}

For POK we find similar results in Figs.~\ref{fig:ccpibrat1} 
and~\ref{fig:ccpibrat5}. The final degree and the final number of 
messages also scale with an exponent close to $0.75$, 
although for sending messages 
there exist some deviations of the most active members 
[Fig.~\ref{fig:ccpibrat1}(a)]. 
Also, the correlations of sending and receiving are linear, 
the same holds for in- and out-degree. 
However, the most active members again deviate with low receiving part, 
i.e. both low number of received messages as well as low in-degree, 
Fig.~\ref{fig:ccpibrat5}. 
Nevertheless, the results for both data sets are mainly consistent and 
the power-law relation Eq.~(\ref{eq:lambda})
is a remarkable regularity.

\subsubsection{Activity and degree distributions}
\label{ssubsec:distributions}

\begin{figure}
\begin{centering}
\includegraphics[width=\columnwidth]{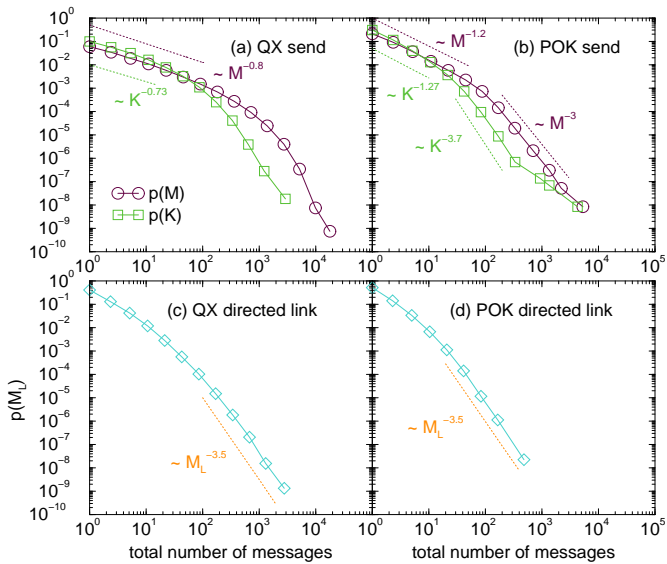}
\caption{
\label{fig:distributions}
Probability densities of activities and degrees. 
The probabilities are plotted versus the total number of messages, $M$, and 
the final degree, $K$, for (a) sending in QX and (b) sending in POK.
The panel (c) and (d) exhibit the probability densities of the total number 
of messages along directed links, $M_L$, for QX and POK, respectively.
The dotted lines serve as guides to the eye and have the indicated slopes.
}
\end{centering}
\end{figure}

Finally, we want to briefly discuss the distributions of activities 
and degrees.
If we assume $p(M)\sim M^{-\gamma_M}$ and $p(K)\sim K^{-\gamma_K}$, then 
with Eq.~(\ref{eq:lambda}) the exponents should be related according to
\begin{equation}
\label{eq:gammalambda}
\gamma_K=1+(\gamma_M-1)/\lambda
\enspace .
\end{equation}

Figure~\ref{fig:distributions}(a+b) displays the probability densities, 
$p(M)$ and $p(K)$, for both online communities. 
Although the distributions are rather broad they do not exhibit straight 
lines in double logarithmic representation.
In panel~(b) we include some guides to the eye with slopes according 
to Eq.~(\ref{eq:gammalambda}) and which roughly follow the obtained curves.
However, $\lambda\approx 0.75$ is relatively close to $1$ so that 
the differences are minor.

The probability densities of activity along direct links are displayed in 
Fig.~\ref{fig:distributions}(c+d) for QX and POK, respectively.
In both cases the frequency of large activity decays approximately 
following a power-law with exponent around $3.5$.

\section{Conclusions}
\label{sec:conclusions}

\begin{table*}
\begin{tabular}{|l||c|c|c|c|c|}
\hline
sending & $H$ & $H_{\rm L}$ & $\beta$ & $\beta_k$ & $\beta_\times$ \\
\hline 
Sec. or Ref. & \cite{RybskiBHLM2009}, \ref{subsec:resltc}, \ref{ssubsec:cascpoisproc} & \ref{subsec:resltc} & \cite{RybskiBHLM2009} & \cite{RybskiBHLM2009} & \ref{ssubsec:resmutgrwmes} \\
\hline\hline
QX & $0.75 \pm 0.05$ & $\approx 0.74$ & $0.22 \pm 0.01$ & $0.22 \pm 0.02$ & $\approx 0.3$ \\
\hline
QX shuffled & $1/2$ &  & $1/2$ &  & $1/2$ \\
\hline
POK & $0.91 \pm 0.04$ &  & $0.17 \pm 0.03$ & $0.17 \pm 0.08$ & \\
\hline
POK shuffled & $1/2$ &  & $1/2$ &  & \\
\hline
BA model &  &  &  & $1/2$ & \\
\hline
CP process & $\rightarrow 1/2$ &  &  &  & \\
\hline
\end{tabular}
\caption{Overview of the obtained exponents. 
QX and POK are the two data-sets. 
For the BA model and the CP process see 
\cite{BarabasiA1999} and \cite{MalmgrenSMA2008}, 
respectively. 
$H_{\rm L}$ is the fluctuation exponent along directed links, 
$\beta_k$ is the growth fluctuation exponent when the degree is considered, and 
$\beta_\times$ is the mutual growth fluctuation exponent based on the 
growth between pairs.}
\label{tab:expos}
\end{table*}

Our work reviews and further supports previous empirical findings 
\cite{RybskiBHLM2009} extending them by some features.
The obtained exponents are summarized in Table.~\ref{tab:expos}.

In addition to \cite{RybskiBHLM2009}, we find very similar 
characteristics for the passivity of receiving as for the 
activity of sending messages. 
This is in line with the strong correlations between individual 
sending and receiving, i.e. most of the messages are somehow replied 
sooner or later. 
Furthermore already the communication 
between two individuals comprises long-term persistence. 

Investigating the probability densities of logarithmic growth rates 
(i.e. growth of the cumulative number of messages between two time steps 
of any member), 
we are able to collapse the curves by scaling them with 
conditional average growth rates and conditional standard deviations.
While less active members follow well the exponentially decaying probability 
density, for the more active members deviations are found in the case of 
large growth rates.

Moreover, we introduce a new growth rate, namely the mutual growth in
the number of messages. 
This is the difference in the number of messages sent between pairs of 
members at two time steps. 
The conditional standard deviation of this mutual growth rate also 
decays as a power-law with increasing initial difference, 
whereas the exponent is close to~$0.3$ and changes to~$1/2$ 
when the data is shuffled. 
We conjecture that this growth reflects cross-correlations in
the activity.

Finally, we propose simulations to reproduce the long-term
correlations and growth properties. 
Basically it consists of generating long-term correlated sequences and 
defining a threshold. 
All values of such sequences above the threshold (POT) represent a 
message event. 
We show that then the correlation and growth features, 
being determined by the imposed fluctuation exponent, 
confirm the relation $\beta=1-H$ \cite{RybskiBHLM2009}.
Including further features, this approach could be a starting point for 
more elaborated modeling of human dynamics.

We would like to note that -- 
except Sec.~\ref{ssubsec:resmutgrwmes} about 
mutual growth in the number of messages 
(and Sec.~\ref{subsec:resoth}) -- 
all analysis and results refer to auto-correlations.
As phenomena, auto- and cross-correlations can occur independently.
However, since most of the messages are replied, 
it is very likely that there are also cross-correlations 
between the members activity, 
which to our knowledge has not yet been studied 
systematically.

Thus, our work opens perspectives for further research activities. 
In particular, the origin of the long-term persistence in the 
communication remains an important question. 
In \cite{RybskiBHLM5a} we demonstrate the relation of $\beta$, $H$ 
with inter-event time scaling. 
From a psychological/sociological point of view one may argue 
where the persistence is originated. 
Is it purely due to a state of mind, solipsistic, emerging from moods, 
or is it due to social effects, i.e. that the dynamics in the 
social network induces persistent fluctuations?
One hypothesis could be that already the social network is 
correlated \cite{RybskiRK2010}.

\section*{Acknowledgments}
We thank C. Briscoe, J.F. Eichner, L.K. Gallos, and H.D. Rozenfeld for
useful discussions.  This work was supported by National Science
Foundation Grant NSF-SES-0624116 and NSF-EF-0827508.
F.L. acknowledges financial support from The Swedish Bank Tercentenary
Foundation.  S.H. thanks the European EPIWORK project, the Israel
Science Foundation, the ONR and the DTRA for financial support.

\bibliographystyle{epj}
\bibliography{gibratmesslong.bib}

\end{document}